\documentclass[aps,prd,superscriptaddress,amsmath,amssymb,preprintnumbers]{revtex4}
\usepackage{graphicx,epsf}
\usepackage{amsfonts}
\usepackage{amssymb}
\usepackage{ulem}
\def\gsim{\mathrel{\vcenter{\hbox{$>$}\nointerlineskip\hbox{$\sim$}}}}
\usepackage[dvips]{color}
\definecolor{red}{rgb}{1,0,0}
\definecolor{green}{rgb}{0,1,0}

\newcommand{\be}{\begin{eqnarray}}
\newcommand{\ee}{\end{eqnarray}}

\begin{document}

\title{Baryon asymmetry of the universe and new neutrino states}

\author{Sebastian Hollenberg}
\email{sebastian.hollenberg@tu-dortmund.de}
\author{Heinrich P\"as}
\email{heinrich.paes@tu-dortmund.de}
\author{Dario Schalla}
\email{dario.schalla@tu-dortmund.de}
\affiliation{Fakult\"at f\"ur Physik, Technische Universit\"at Dortmund, D-44221 Dortmund, Germany}

\begin{abstract}
The presence of additional neutrino states with masses in the GeV range is allowed by electro-weak precision observables.
However, these additional states can lead to lepton number violating
interactions which potentially can wash out considerably
any GUT scale generated or elsewise pre-existing baryon asymmetry. 
We discuss the resulting bounds on neutrino parameters and find that - unless
the baryon asymmetry is created at or below the electroweak scale or in some
flavor which is decoupled from interactions to the new neutrino states -
these
have to be pseudo-Dirac with
mass splittings between the right-handed and left-handed states of less
than about ten keV.     
\end{abstract}
%
%
\preprint{DO-TH-11/22}

\maketitle

Recently, the possible existence of new neutrino states beyond the known three light
neutrinos has gained considerable interest, both in models with a fourth fermion
generation \cite{Burdman:2009ih,Carpenter:2010sm,Lenz:2011gd,Hou:2010qx} as well as a possible solution to neutrino oscillations anomalies
observed in solar and reactor neutrino oscillations and the LSND and  MiniBooNE
experiments \cite{Barry:2011wb,deHolanda:2010am,Kopp:2011qd}, as well as a candidate for warm dark matter 
\cite{Keung:2011zc, Asaka:2005an}
and due to other attractive
astrophysical properties \cite{Kusenko:2006rh}.

In principle such new neutrino states can be sterile $SU(2)$ singlets or $SU(2)$ doublets and
Majorana as well as Dirac particles. In the most general case (and according to the
theoretical prejudice) neutrinos are Majorana particles which possess both Dirac and Majorana masses. 

Majorana particles can, for example in various versions of the seesaw mechanism,
explain why the three known neutrinos are so much lighter than the charged leptons 
and quarks. On the other hand, neutrino Majorana masses violate lepton number which
induces baryon number violation via sphaleron transitions in the early universe.
Consequently Majorana neutrinos can contribute to both the generation 
of the baryon asymmetry in the
universe via leptogenesis, as well as to the washout of any GUT scale generated or
elsewise pre-existing  lepton or baryon -- or more accurately 
baryon minus lepton $B-L_{\alpha}$ asymmetry --
via their lepton number violating interactions \cite{Blanchet:2009kk,Murayama:2010xb}.
On the other hand, for Majorana neutrinos with masses in the GeV range, any GUT scale generated baryon 
asymmetry cannot be protected by the
effective freeze out of heavy neutrino states 
roughly at their mass scales. On the contrary baryon or lepton number violating interactions
emerging from the presence of additional Majorana states threaten the survival of the asymmetry, which has
to bridge the gap of some twelve orders of magnitude down to the electro-weak symmetry 
breaking scale $T_{\rm EW}$
(for related works, see e.g. \cite{Fukugita:1990gb,KlapdorKleingrothaus:1999bd,KlapdorKleingrothaus:2000dm}).

The introduction of a Majorana mass $M_R$ in presence of a Dirac mass $m_D$ leads to two mass eigenstates which, in the seesaw 
limit $m_D \ll M_R$, decouple a light (active) and a heavy (sterile) Majorana neutrino.
However, it has been shown that this limit can only be valid for the established first three neutrino families. 
Additional neutrinos such as a fourth generation are strongly constrained from neutrinoless double beta
decay to have small Majorana masses, as long as they mix with the electron neutrino
via a non-vanishing PMNS matrix element $U_{e4}$~\cite{Lenz:2011gd}. 
Because their Dirac masses $m_D = y v$
cannot be larger than TeV scale (assuming the Yukawa coupling $y$ to be
maximally of order one) and due to the fact that their maximal mass difference to their charged lepton partner has been found
to be less than 140 GeV from electro-weak precision measurements \cite{Eberhardt:2010bm}, 
their Majorana mass has to be smaller than $  \mathcal O(10 ~\rm GeV)$ to
agree with the non-observation of neutrinoless double beta decay
(in a strict seesaw model, even
stronger bounds may be obtained by using unitarity relations \cite{Blennow:2010th}. For 
complementary bounds obtained from applying absolute neutrino mass bounds on
light neutrino species 
radiative mass contributions induced by the new neutrino states see \cite{Aparici:2011nu,Aparici:2011qu}). This
limit, $m_D \gg M_R$, is called pseudo-Dirac limit. The resulting mass
eigenstates are almost degenerate with a small mass splitting $\delta m =
M_R$ and can be treated as one object as they do not decouple into
active and sterile states. Lepton number violating processes triggered by
$M_R$ are therefore suppressed. In the limit
of vanishing mass splitting $\delta m = M_R \rightarrow 0$ the pseudo-Dirac
neutrino becomes a pure Dirac neutrino and such processes are forbidden.
Note, that additional $SU(2)$ doublet neutrinos are also tightly constrained from the decay width of the 
$Z$ boson~\cite{:2005ema}, resulting in a lower mass bound of $m_\nu>m_Z/2$.

In this letter we study the cosmological consequences of such additional species of pseudo-Dirac neutrino states with masses at an 
intermediate scale of 
some $\mathcal O(100 ~\rm GeV)$. Above the electro-weak phase transition around some
$T_{\rm EW} = 246 ~\rm GeV$ the Dirac mass of the additional neutrino species vanishes due to its proportionality to the
Higgs vacuum expectation value via their Yukawa couplings; the Majorana mass of the states, however, shows no such behavior.
Therefore, the mass splitting $\delta m$ between the left-handed and right-handed neutrino states is not temperature-dependent.
It is important to keep in mind here that for the pseudo-Dirac neutrino the usual seesaw
mechanism does not apply: the Majorana mass is {\it not} assumed to be much larger than the Dirac mass of the neutrino but rather
small as, e.g., $M_{R} \sim \mathcal O(100 ~\rm GeV)$. 
Thus at temperatures above the electro-weak phase transition pseudo-Dirac neutrinos behave exactly as light sterile neutrinos 
bearing only a Majorana mass. As a consequence our estimates provide also constraints on such neutrinos. 
However, we will focus on additional $SU(2)$ doublet neutrinos with a right-handed partner
in the pseudo-Dirac limit.

Such additional $SU(2)$ doublet neutrinos give rise to lepton number violating scattering processes such as the scattering
of lepton doublets $l^{}_{L}$ to Higgs particles $\phi$ via, e.g., $l^{}_{L} \phi \rightleftharpoons l_{L}^{\rm c} \phi^{\rm c}$ mediated by
the additional 
pseudo-Dirac-type neutrinos above the electro-weak phase transition. The lepton number violation entailed in such processes is
then converted into baryon number violation via sphalerons. Hence, any lepton number violating scattering processes above the 
electro-weak phase transition endanger any pre-existing lepton or baryon asymmetry in the epoch between the GUT scale and the 
electro-weak scale as long as they are in thermal equilibrium.

However, the observation of a non-vanishing baryon asymmetry suggests that such depletion should not have taken place; put another
way, the requirement that lepton number violating processes mediated by additional neutrino species are weak enough as not to 
dilute away a pre-existing lepton (or baryon) asymmetry implies that such processes should not come into 
thermal equilibrium until the electro-weak phase transition. The bound derived thus provides a conservative estimate for the parameter space allowed.

We can now derive bounds on the masses of the additional neutrino states by invoking the out-of-equilibrium criterion. To this end, 
our first objective is parameterizing the thermally-averaged scattering cross section. Obviously, since we are not dealing with a
temperature regime any more in which the approximation $z \equiv \delta m / T \gg 1$ is still valid (as would be the case for the GUT scale 
neutrino states), we find that
\be
   \langle \sigma |v| \rangle \sim 
   \sum_{\beta,\gamma} y_{\alpha\beta}^2 y_{\alpha\gamma}^2 U_{\beta 4}^2 U_{\gamma 4}^2 
   \frac{\delta m^2}{\left(T^2 + \delta m^2\right)^2} 
   =\sum_{\beta,\gamma} y_{\alpha\beta}^2 y_{\alpha\gamma}^2 U_{\beta 4}^2 U_{\gamma 4}^2 
   \frac{1}{\delta m^2}\frac{z^4}{\left(1+z^2\right)^2},
\ee
presents a reasonable approximation \cite{Kolb:1990vq}.
We have assumed that the lepton asymmetry has been generated originally in the
flavor $\alpha$; $y_{\alpha \beta}$ denotes the Yukawa couplings in flavor space, whereas $U_{\alpha\beta}$ is the neutrino mixing matrix.    
It is readily seen that in the limit $z \gg 1$ we recover
the thermally-averaged cross section for GUT scale neutrino states (given by $\langle \sigma |v| \rangle \sim 1/\delta m^2$). 
Furthermore,
we infer from this parameterization that as the Majorana mass goes to zero, so does the scattering cross section; this paraphrases
the fact that for pure Dirac states there is no washout of the lepton asymmetry expected.

With the appropriate scattering cross section at hand, we can now write down the relevant rate for the washout process,
\be
 \Gamma(T) \sim n_{\phi} \langle \sigma |v| \rangle.
\ee
Here $\Gamma$ contains the number density $n_{\phi} \simeq 2 \zeta(3) T^3 / \pi^2$ of Higgses in which $\zeta$ is the 
Riemann zeta function. 

The washout of lepton number $L$ between GUT scale and
electro-weak scale can then be calculated straightforwardly via
\be
 \ln\left(\frac{L(z_{\rm EW})}{L(z_{\rm GUT})}\right) = 
 - \frac{1}{H(z=1)} \int_{z_{\rm GUT}}^{z_{\rm EW}} \mathrm d z' ~z' n_{\gamma} \langle \sigma |v| \rangle(z'),
\ee
where $n_{\gamma} \simeq n_{\phi}$ is the photon number density.
Here
\be
H(T)= \sqrt{\frac{8\pi^3g_{\ast}}{90}} \frac{T^2}{m_{\rm Pl}}
\ee 
is the Hubble rate, $m_{\rm Pl}$ the Planck mass, and $g_{\ast}$ the effective number of degrees of freedom. 

\begin{figure}
 \centering 
 \includegraphics[scale=1.5]{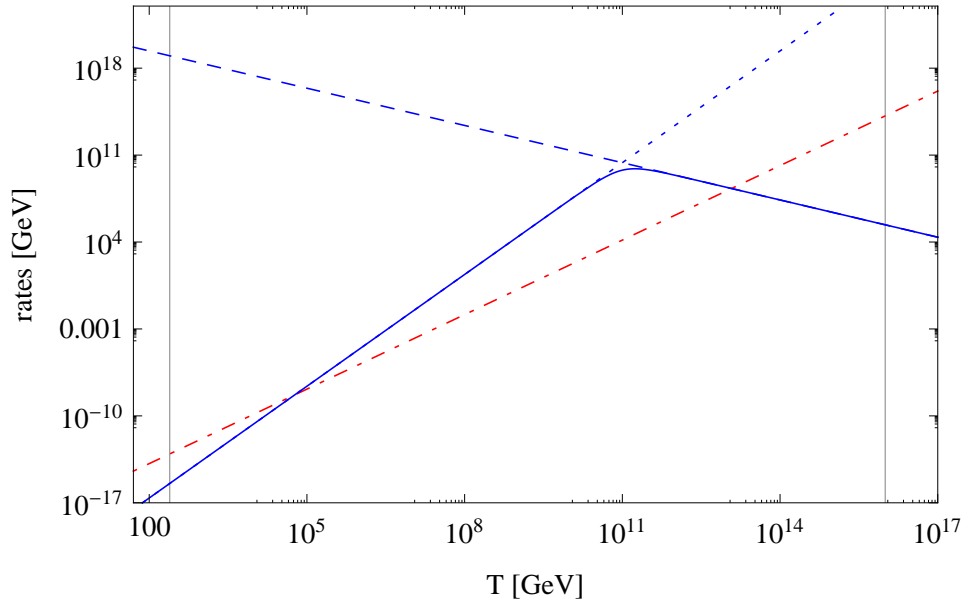}
 \caption{An illustrative log-plot of the rates involved in the freeze out of heavy additional Majorana neutrino states as a function
          of temperature. 
          The dashed-dotted line gives the Hubble rate $H(T)$, the solid line represents the scattering rate $\Gamma(T)$ alongside
          its asymptotic cases; the dotted line gives the case of $T \ll \delta m$, whereas the dashed line shows the 
          reciprocal limit of $T \gg \delta m$. The vertical line to the left marks the electro-weak threshold 
          $T_{\rm EW} = 246 ~\rm GeV$, the vertical line to the right marks the GUT threshold $T_{\rm GUT} = 10^{16}~ \rm GeV$.
          For illustrative purposes a Majorana mass of $\delta m = 10^{11} ~\rm GeV$ has been chosen. It is seen that the scattering
          rate exceeds the expansion rate in different temperature regimes as well as that the expansion rate dominates in others.
          Note, also, that the intersections between the Hubble rate and the scattering rate are given by the intersections of the
          asymptotes of the scattering rate to a good approximation. 
          We give a detailed account of the physics of the entailed washout processes in the text.}
 \label{rates}
\end{figure}

\begin{figure}
 \centering 
 \includegraphics[scale=1.5]{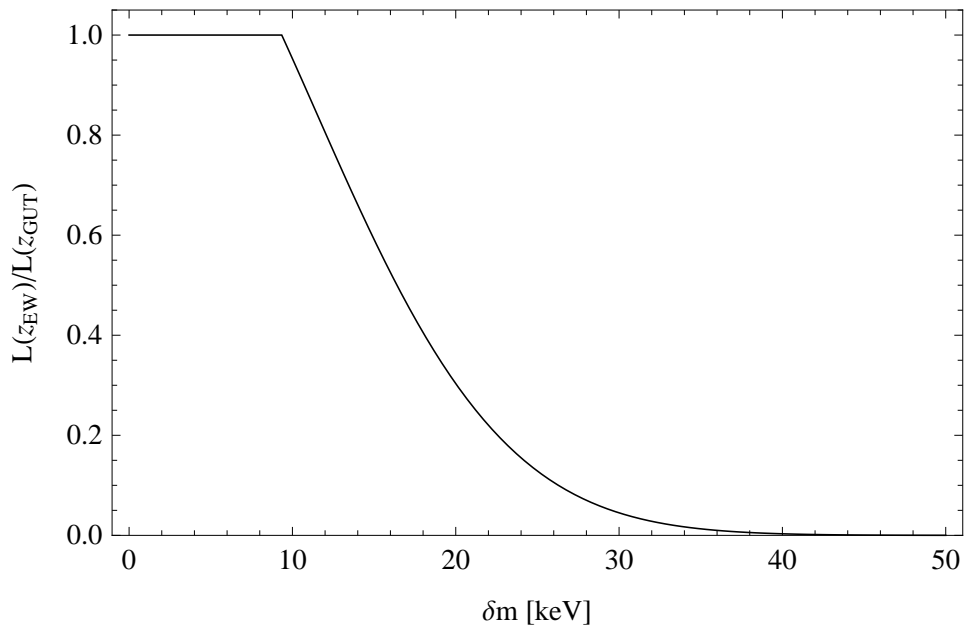}
 \caption{The ratio of the lepton asymmetry at the electro-weak scale and the GUT scale as a function of the Majorana mass. Yukawa couplings and mixing elements have been chosen to $\sum_{\beta,\gamma} \left( y_{\alpha\beta} y_{\alpha\gamma} U_{\beta 4} U_{\gamma 4} \right)^{-2} = 1$.
          It is seen that up to some $\delta m \sim 10 ~\rm keV$ there is no washout of the pre-existing lepton asymmetry at all,
          which can be traced back to the fact that for such Majorana masses the scattering rate simply cannot exceed the Hubble rate.
          Beyond this mass bound scattering channels open up as the scattering rate begins to exceed the Hubble rate; 
          washout processes inevitably deplete any pre-existing lepton asymmetry with increasing values of the Majorana mass.}
 \label{washout}
\end{figure}

There is, however, a minor caveat concerning the integration limits in the integral. 
The washout processes will be active only
during thermal equilibrium, $\Gamma \gsim H(T)$.
Depending on the Majorana mass of the neutrino, 
the scattering rate can fall below the Hubble rate for temperatures both below the GUT and above the electro-weak scale; the
lepton number violating processes would thus fall out of thermal equilibrium rendering the washout processes impotent. 
This situation is illustrated in FIG.\ref{rates}, where for a certain choice of parameters it can be seen that the reaction
falls out of equilibrium both considerably above the electro-weak scale and below the GUT scale already. In all such cases 
we choose the intersection points of the Hubble rate and the scattering rate as integration limits and set the lepton number
washout to zero otherwise as to not to overestimate the lepton number depletion. We then solve the integral with those
integration limits. The result is shown in FIG.\ref{washout}. For certain masses the scattering cannot come into thermal 
equilibrium until the electro-weak scale and hence there cannot be any depletion of pre-existing lepton number. This is the 
case for Majorana masses less than 
\be
 \left(\frac{\delta m}{10 ~{\rm keV}}\right)^2 \lesssim 1 \times 
 \sum_{\beta,\gamma} \left(\frac{1}{y_{\alpha\beta} y_{\alpha\gamma} U_{\beta 4} U_{\gamma 4} }\right)^2 
 \times \left(\frac{T_{\rm EW}}{246 ~{\rm GeV}}\right)^3.
\ee
Beyond this threshold lepton number violating scatterings
can come into equilibrium and pre-existing lepton number gets diluted away exponentially as the neutrino mass increases.
If, for instance, a drop in lepton number by one order of magnitude is considered compatible, the bound on the Majorana mass 
is relaxed to some $\delta m \sim 25 ~\rm keV$.


This bound of the pseudo-Dirac mass splitting improves bounds previously obtained from neutrinoless 
double beta decay searches by at least three orders of magnitude. 
It has to be understood as a rough estimate, though, as it is subject to the exact values of Yukawa couplings and PMNS matrix elements 
describing the mixing of the flavor(s) in which the baryon asymmetry has been generated
and the new neutrino states.
Taking this into account
the results obtained in this letter apply for any additional 
Majorana neutrino species and constrain their Majorana mass to be in the keV range. 
This, in turn, implies that any additional 
species has to be of a pseudo-Dirac type by virtue of its small Majorana mass
unless the baryon asymmetry is generated at or below the electroweak scale and/or flavor protected. Note, also, that the analysis 
presented in this
letter thus independently confirms earlier findings in the context of a complete sequential fermion generation 
\cite{Lenz:2011gd}.

In summary we have discussed the washout of a pre-existing or GUT scale generated baryon 
asymmetry due to the presence of new neutrino states.
We find that pure Majorana neutrinos are limited to masses below the order of ten keV.
Such light neutrino states have to be $SU(2)$ singlets to avoid the bound from the
Z boson decay width.
Weak scale pseudo-Dirac neutrinos are allowed only with tiny mass splittings of the order of ten keV. 
Thus a fourth generation neutrino being affected by the Z boson width bound 
has to be Dirac or pseudo-Dirac with tiny lepton number violation.

\acknowledgments
We thank A. Kartavtsev for useful discussions during the course of this work. SH wants to thank the Saha Institute of
Nuclear Physics, Kolkata and in particular P.B. Pal for hospitality during the final stage of this work.

\end{document}